\documentclass[onecolumn, a4paper, 10pt]{IEEEtran} 
\usepackage{graphics}
\usepackage{amsmath}
\usepackage{url}
\makeatletter

\usepackage{epsfig}
\usepackage{subfigure}
\usepackage{graphicx}
\usepackage{longtable}
\usepackage{algorithm}
\usepackage{algorithmic}
\usepackage{times}
\usepackage{verbatim}
\def\unnumfootnote{\xdef\@thefnmark{}\@footnotetext}

\makeatother
\begin{document}%
\title{An estimation of cattle movement parameters in the Central States of the US}
\author{\itshape{Phillip Schumm}\footnotemark${^{1}}$$^{,*}$, \itshape{Caterina Scoglio}\footnotemark${^{1}}$$^{,*}$, \itshape{H. Morgan Scott}\footnotemark${^{2}}$}
\maketitle
\thispagestyle{empty} 
\footnotetext[1]{Kansas State University, Electrical and Computer Engineering Department, Sunflower Networking Group}
\footnotetext[2]{Kansas State University, Department of Diagnostic Medicine/Pathobiology}
\footnotetext[3]{(*)Corresponding authors: {pbschumm, caterina} at ksu.edu}

\unnumfootnote{}
\begin{abstract}
The characterization of cattle demographics and especially movements is an essential component in the modeling of dynamics in cattle systems, yet for cattle systems of the United States (US), this is missing. Through a large-scale maximum entropy optimization formulation, we estimate cattle movement parameters to characterize the movements of cattle across $10$ Central States and $1034$ counties of the United States. Inputs to the estimation problem are taken from the United States Department of Agriculture National Agricultural Statistics Service database and are pre-processed in a pair of tightly constrained optimization problems to recover non-disclosed elements of data. We compare stochastic subpopulation-based movements generated from the estimated parameters to operation-based movements published by the United States Department of Agriculture. For future Census of Agriculture distributions, we propose a series of questions that enable improvements for our method without compromising the privacy of cattle operations. Our novel method to estimate cattle movements across large US regions characterizes county-level stratified subpopulations of cattle for data-driven livestock modeling. Our estimated movement parameters suggest a significant risk level for US cattle systems.
\end{abstract}

\section{Introduction} 
Livestock systems serve significant roles for many regions across the world, yet past outbreaks of disease have demonstrated that they can possess a number of vulnerabilities \cite{BMP2003, FDA2001, KWSMCHCKWG2001, GKK2006, KGJK2007, BBSC2012}. The livestock systems of the United States (US), though strictly regulated, may yet be found susceptible to foreign diseases such as Rift Valley Fever \cite{XCSS2013}. The successful modeling and analysis of livestock epidemics for any region relies heavily on an understanding of the underlying system components. The three most critical elements in a practical epidemic model are the disease progression model, the geo-spatial characterization of the susceptible populations, and the spatial-temporal description of the interactions of individuals within the system \cite{CBBVV2007, CV2008, BCGHRV2009}. The models of disease progression are several and often independent of the region studied \cite{MA1979, PV2001, BBV2008, V2012, DSV2013}. Data-driven, spatial characterizations of populations are available through regularly conducted censuses (censi) \cite{US_CENSUS_2013, US_AGCENSUS_2013}. The third element, the interactions of individuals within the system, represents the set of spatial movements of individuals. When considering system-wide outbreaks of disease, the impact of movement parameters has been shown to be as significant as that of epidemic parameters in metapopulation models \cite{CV2007, BV2012}. Domestic livestock systems fit well in such metapopulation models as the movements of livestock are controlled and the individuals are restricted to reside within populations rarely defined by their choice. Within the US, livestock movements are controlled by the cattle industries, primarily beef, dairy, breeding, and showmanship.\\

Within Europe, motivated by outbreaks of Foot and Mouth Disease, a number of governments have designed and implemented animal tracking systems even to the resolution of individuals' daily movements. The databases created by these studies have generated very detailed characterizations of livestock movements for a number of European nations \cite{GKK2006, KGJK2007, BBSC2012, EC2000, W2005, KDGK2006, BTCMG2006, KGK2006, BBMG2007, REC2007, RC2007, BNAL2008, BKC2008, VK2009, NGSPPFC2009, MAF2009, KDVH2010, VHSW2010, RDD2011}. No similar program has yet to be implemented for the United States, although some have long been in preparation \cite{NAIS2013}. In the US, a cultural appreciation of personal privacy from the government, strong competition between meat production companies, and a U.S. Federal privacy protection law restrict the ability of the government to collect and release livestock data at a finer spatial resolution than is currently done through the United States Department of Agriculture's (USDA) Census of Agriculture \cite{US_AGCENSUS_2013}. To address this challenge, a number of survey-based methods have been used to study livestock movements across small regions \cite{BBT1995, FHDW1998, BTC2001, SM2007, D2007, THBCOASK2009}. However, the national scale of US cattle trade and the potential for livestock diseases to impact the entire country necessitate movement data, models, or estimates to be determined for larger regions. Recently a study has been published of a nation-wide movement estimation based on a $10\%$ sample of veterinary records from State border-crossing cattle shipments \cite{BGPMLW2013}. This impressive study, although the first of its magnitude, only captured shipments of cattle that crossed state borders. Although it offers a picture of state-to-state shipment counts, the method used is unable to capture the livestock movements within each state.\\

In this paper, we formulate a large, convex optimization problem to estimate parameters describing the movements of cattle within $10$ Central States of the United States. We collect cattle population and aggregated movement data from the United States Department of Agriculture's database and optimally estimate anonymous data points to construct a database of inputs for an estimation of cattle movement parameters. We design the estimation method to produce a high resolution of cattle demographic and movement parameters and to include the minimal set of assumptions. Our results produce county-to-county movement probabilities among stratified subpopulations as well as birth, slaughter, and expiration rates of cattle for $1034$ US counties. In section \ref{datasection}, we describe the USDA data structures and challenges present in the database. We estimate non-disclosed data points and discuss the mapping of USDA data to inputs for our estimation formulation. In section \ref{formulation}, we formulate the estimation problem and describe the maximum-entropy objective and the flexible set of linear constraints with parameters sculpted to the USDA data set and as few assumptions as possible. We solve the optimal estimation problem and display a subset of the results in section \ref{resultssec}. Section \ref{concl} summarizes this paper with a discussion of the results, a series of questions for future agricultural census distributions, and a calculation of critical movement thresholds that demonstrate the potential for a disease outbreak in the cattle of this region.

\section{Data collection and structure} \label{datasection}

\begin{figure}
	\includegraphics[width = 3.2in]{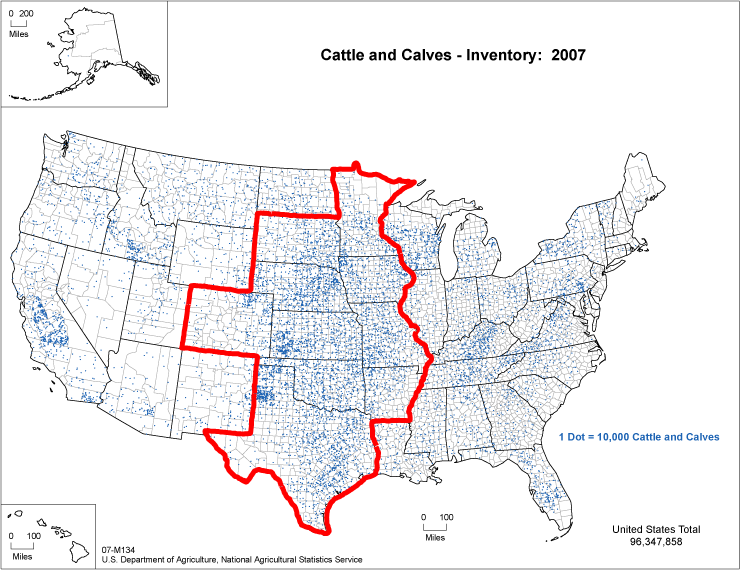}
	\caption[The $10$ Central States]{The $10$ Central States of interest are outlined with a red trace over the population distribution of cattle in the United States as provided from the United States Department of Agriculture \cite{NASS2013}. Each blue dot represents $10,000$ head of cattle.}
	\label{USDAoutline}
\end{figure}

Every five years, the United States Department of Agriculture (USDA) conducts the United States Census of Agriculture \cite{US_AGCENSUS_2013}. The National Agricultural Statistics Service (NASS) of USDA then summarizes and publishes a large set of data covering livestock, crops, operator demographics, and much more \cite{NASS2013}. As the most comprehensive and clean database of US livestock statistics, the Census of Agriculture as presented in the NASS database is used for our estimation of cattle movement parameters. In particular, we use data from the $2007$ Agricultural Census as the $2012$ data was not published at the time of this study. The data of interest to this work comes from section $13$, titled ``Cattle and Calves'', on page $10$ of the $2007$ Agricultural Census. Section $13$ also has a set of related instructions located on page $2$ of the instruction sheet appended to the Agricultural Census \cite{US_AGCENSUS_2013}. From the U.S. Census Bureau and their $2010$ Census (of humans) in the United States, we use the centers of human population for each county \cite{US_CENSUS_2013}. We include these geographical points to consider a basic quantification of distance for the cattle movement estimation. Adding this geography to the data from the NASS database, we estimate sets of parameters to characterize cattle movements in the States of Arkansas, Colorado, Iowa, Kansas, Minnesota, Missouri, Nebraska, Oklahoma, South Dakota, and Texas \cite{LGHDRF1997, FU2003}. The US beef production feedlot structure produces more frequent and larger flows of cattle than the typical grazing structure, and notably, these $10$ Central States form the core of the US feedlot industry \cite{FHDW1998, BGPMLW2013}. As outlined in figure \ref{USDAoutline}, these States contain $1034$ US counties with more than $51$ million head of cattle of the $96.3$ million head reported in the $2007$ Agricultural Census \cite{NASS2013}.

\subsection{Data structure of USDA NASS}

From section $13$ of the $2007$ Agricultural Census, we are primarily interested in the responses to questions concerning the populations and movements of cattle. For cattle populations, the Agricultural Census identifies the total number of cattle (Question $3$), the number of dairy cows kept for production of milk (Question $2$.b), and the number of cattle, including calves, who were in a preslaughter feed program (Question $5$) on December $31$, $2007$. For the movements of cattle, composed of all sales and shipments, the Agricultural Census captures the total number of cattle ``sold or moved'' during $2007$ (Question $4$) and the total number of cattle shipped directly to a slaughter market from a preslaughter feed program during $2007$ (Question $6$) \cite{US_AGCENSUS_2013}. Although these data are collected for each individual operation, the statistics of the cattle populations are reported only through aggregated distributions that are delineated by the operation's county, the type of cattle, and the number cattle of a particular type. The sizes of populations are sorted into seven standard ranges: $1-9$ cattle, $10-19$ cattle, $20-49$ cattle, $50-99$ cattle, $100-199$ cattle, $200-499$ cattle, and $500$ or more cattle. The statistics of the cattle movements sort the responses by the total yearly movements (or slaughter) across the same $7$ standard ranges, the type of cattle, and the county of the originating operation. It is worth noting that the total dairy cattle population and the total preslaughter population of a given county are subpopulations of the total cattle population for that county. Similarly, the total number of cattle shipped to slaughter from a preslaughter feed program is a fraction of the total cattle movements (sales and shipments) for each county.\\

According to appendix A of the $2007$ Agricultural Census Summary and State Data report, the data presented in the NASS database has undergone some initial processing and systematic error correction \cite{VC2009}. This results in a very consistent database and the potential errors induced by these methods have been quantified in the same appendix. Even with these diligent efforts, there remain two significant challenges in utilizing the data to characterize the movements of cattle. The first concerns the resolution of the timescale of the data.
As a summary of the entire year $2007$, these data fail to capture any seasonal fluctuations in the cattle populations and movements \cite{DDM2004}. This challenge arises from the administration process of the Agricultural Census and we acknowledge its significance; however, we find no comprehensive, data-driven solution to the seasonality challenge and consider only mean-field probabilities in our estimation. The second challenge is posed by the direct sorting of the census responses into the $7$ standard ranges rather than preserving any operation-based connections between data points. Therefore, a population of $50$ dairy cattle might belong to any operation having a total number of cattle greater than or equal to $50$ ($4$ possible size ranges) without having any connection to the size of its entire operation. Similarly, the sizes of shipments have no direct connections to the size of the originating operations besides a few loose feasibility restrictions.\\

The data, as it comes from NASS, has been released in such a way that the information of individual farms and cattle operations is not revealed. This is done intentionally by USDA to comply with Title 7 of the U.S. Code \cite{VC2009}. To maintain this anonymity, critically selected elements of the data have not been disclosed. We will estimate these non-disclosed data points through a pair of tightly-constrained convex problems and then include them in the inputs for our main problem, the estimation of cattle movement parameters across the $10$ Central States.

\subsection{Data structure for estimation problem}

To estimate the non-disclosed entries in the original data, we construct a pair of optimization problems, one for the population data and a second for the movement and slaughter data. The objective of both formulations is a maximum entropy function. For the population distributions and by each State, we maximize the entropy of the distributions of each cattle type given by $Type_{A} = \{Dairy$, $Preslaughter$, $All$ $Cattle\}$, where the distributions are normalized by their respective State totals. Likewise, the entropy of the normalized distributions describing the totals of cattle slaughter shipments and cattle movements is maximized across shipment type, $ShipType = \{ All$ $Shipments$, $Slaughter\}$, shipment size, $Size_{A}=\{z1\_9$, $z10\_19$, $z20\_49$, $z50\_99$, $z100\_199$, $z200\_499$, $z500\_up\}$, and county. We present the formulations of these problems in appendix A. These problems are solved for each of the $10$ States, filling in all non-disclosed data entries. We quantify the dependence of our inputs on these estimations with both the fraction of cattle and the fraction of populations estimated for each State in Appendix B.\\

We would prefer to represent the system of cattle through three subpopulations rather than the two subpopulations and total population of $Type_{A}$. The mapping of the first set $Type_{A}$ to a set of three cattle subpopulation types is not trivial as is requires the expression of a relationship among the $3$ cattle types of $Type_{A}$. For the county totals of these three types, the relationship $Tc_{Dairy,c}^{x} + Tc_{Preslaughter,c}^{x} \leq Tc_{All Cattle,c}^{x}$ holds, where $Tc_{\cdot,c}^{x}$ represents the respective county total for county $c$ and $x$ indicates that this is a variable to be estimated. However, the relationship is not guaranteed if we consider the stratification of the populations by size as
\begin{equation}
Pop_{Dairy,c,i} + Pop_{Preslaughter,c,i} \le Pop_{All Cattle,c,i} \text{     } \forall \left(c \in County, i \in Size_{A}\right) \text{ ,}
\label{sizepoprel}
\end{equation}
where $Pop_{\cdot,c,i}$ represents the respective population total for operations with size $i$ and in county $c$. Rather, Broomfield County in the State of Colorado, as do a number of other counties, reports data in violation of this relationship. Broomfield County is a irregularly cut county on the north side of Denver, Colorado and has $2$ dairy farms with total populations in the range $z100\_199$. The county, however, reports $2$ populations of Dairy cattle in the size range $z50\_99$ and no total ($All$ $Cattle$) populations in the size range $z50\_99$. Thus the left-hand side of inequality \ref{sizepoprel} would be non-zero while the right-hand side is identically $0$ for $i = z50\_99$, $c = Broomfield\_Colorado$. This discrepancy arises from other cattle residing at both of these operations that raise the total operation populations into the next size range. We find that through an aggregation of the sizes into $3$ ranges, $Size_{B}=\{z1\_19$, $z20\_199$, $z200\_up\}$, it becomes feasible to assume the relationship of inequality \ref{sizepoprel} for each size $j$ in $Size_{B}$.\\

Let us define a new cattle type $Beef$, representing all cattle that are not serving as dairy cattle nor in a preslaughter feed program, as the difference between the total population $All$ $Cattle$ and the two subpopulation types, $Dairy$ and $Preslaughter$. The $Beef$ type represents a diverse set of cattle operations including grazing, backgrounding, and breeding services. The name $Beef$ is chosen for simplicity with the assumption that this is the majority role served by cattle in this type. At this point, we describe the cattle subpopulations $Pop_{t,c,j}$ by cattle type $t$ in $Type_{B} = \{Dairy, Preslaughter, Beef\}$, county $c$, and size $j$ in $Size_{B}=\{z1\_19$, $z20\_199$, $z200\_up\}$. The set of subpopulations $\{Pop_{t,c,j}^{R}\}$ results from the solution of the first of the data-patching optimization formulations and represents an aggregation of all cattle type-based subpopulations fitting the subpopulation descriptors, yielding only $9$ subpopulations per county. County totals for sales $Tc_{All Movement,c}^{\left(s\right),x}$ and slaughter $Tc_{Slaughter,c}^{\left(s\right),x}$, as well as the distributions of yearly totals of sales $Sales_{All Movement,c,i}^{x}$ and slaughter $Sales_{Slaughter,c,i}^{x}$ stratified by county $c$ and size $i$ in $Size_{A}$ were completed through the solution of the second data-patching optimization problem of appendix A.

\section{Cattle movement parameter estimation} \label{formulation}

We formulate a non-linear, yet convex, optimization problem with an objective to maximize the entropy of the out-going distributions of each subpopulation \cite{Y2008}. The formulation is nearly linear with the exception of the objective function, or equivalently stated, this formulation contains only linear constraints. The choice of maximum entropy for the problem objective aims to predict the solution having the minimal assumptions (maximum uncertainty) beyond the information contained within the set of constraints \cite{Y2008}. We chose this form of objective to not force any artificial objective in our estimation. Nevertheless, assumptions have been made in the design of both the variables and the constraints. We assume that 
\begin{itemize}
  \item There are no outgoing movements from preslaughter feed programs except for the outgoing movements of cattle for slaughter,
	\item Cattle classified as dairy cattle do not move into preslaughter feed programs,
	\item Populations of preslaughter feed cattle having population sizes of $200$ head of cattle or more are responsible for all shipments to slaughter that result in yearly totals of $500$ or more head shipped from a single premise,
	\item All sub-populations remain constant on a year-to-year basis, and
	\item The counties considered form a closed system with no significant movement into or out of the system.
\end{itemize}

\subsection{Problem formulation}

The central portion of this paper revolves around the formulation to estimate the cattle movement parameter $p^{x}_{t_1,j_1,t_2,j_2,dist}$, which is a probability that represents the movement process from an origin subpopulation of type $t_{1}$ and size $j_{1}$ to a destination subpopulation of type $t_{2}$ and size $j_{2}$ with a distance falling in a discrete distance range $dist$ between the origin and destination counties. We mark the decision variables of this formulation with a superscript $x$ to distinguish them from the parameters of the problem. This formulation also estimates the birth $bt^{x}_{c_1,t_1,j_1}$, expiration of utility (cull) $dt^{x}_{c_1,t_1,j_1}$, and slaughter $sl^{x}_{c_1,t_1,j_1}$ probabilities for each sub-population of each county. The three types of cattle are now rearranged into the set $Type_{B}$ as $\{Dairy, Preslaughter, Beef\}$, with size ranges defined by $Size_{B}=\{z1\_19$, $z20\_199$, $z200\_up\}$. The set of discrete distance ranges used in this formulation is called $Distance$ and is defined as $\{d0, d100, d200, d500, d1000, d_{too far}\}$. The number noted in each distance range between two county centers should be read as the maximum distance in miles of the range with the minimum defined by the previous level, for example, $d500$ indicates a distance between the two county centers falling between $200$ and $500$ miles. The closest range, $d0$, is assigned for any pair of counties with centers less than $10$ miles apart, as well as each county with itself. The formulation has an objective to maximize the entropy of the outgoing distributions of all sub-populations as follows.

\begin{IEEEeqnarray}{c}
\text{Maximize   } J \nonumber \\
\text{where   } \nonumber \\
J = \sum_{c_{1}}\sum_{t_{1}}\sum_{j_{1}} \Bigg[ -st^{x}_{c_1,t_1,j_1}\text{log}\left(st^{x}_{c_1,t_1,j_1}\right)
- \text{  } sl^{x}_{c_1,t_1,j_1}\text{log}\left(sl^{x}_{c_1,t_1,j_1}\right)
- \text{  } dt^{x}_{c_1,t_1,j_1}\text{log}\left(dt^{x}_{c_1,t_1,j_1}\right) \nonumber\\
+ \sum_{ c_{2} }\sum_{t_{2}}\sum_{j_{2}} -p^{x}_{t_1,j_1,t_2,j_2,D\left(c_1,c_2\right)}\text{log}\left(p^{x}_{t_1,j_1,t_2,j_2,D\left(c_1,c_2\right)} + 1.0 - f_{t_1,j_1,t_2,j_2,D\left(c_1,c_2\right)}\right) \Bigg] \nonumber\\
\text{and } c_{1} \in County, t_{1} \in Type_{B}, j_{1}\in Size_{B}, \nonumber\\
c_{2} \in \{County \vert D\left(c_1,c_2\right) \neq d_{too far}\}, t_{2} \in Type_{B}, j_{2}\in Size_{B} \text{ .}
\label{PE_Objective}
\end{IEEEeqnarray}

The out-going probability distribution for the cattle of each subpopulation is completed through the inclusion of a probability to remain or stay, $st^{x}_{c_1,t_1,j_1}$, in the origin subpopulation. The sum of the entropy of these distributions composes our objective function. We have implemented industrial constraints with a set of parameters $\{ f_{t_1,j_1,t_2,j_2,dist} \}$ described in the following constraints which forces a subset of the movement probabilities $p^{x}_{t_1,j_1,t_2,j_2,dist}$ to zero. We account for this by introducing a complimentary $\left(1.0 - f_{t_1,j_1,t_2,j_2,dist}\right)$ in the logarithm of $p^{x}_{t_1,j_1,t_2,j_2,dist}$ to avoid the computation of the natural logarithm of zero.

\[
\text{Subject to}
\]

\[
\text{Constraints on Statistical rules}
\]

\begin{equation}
p^{x}_{t_1,j_1,t_2,j_2,dist} \le f_{t_1,j_1,t_2,j_2,dist} \text{     } \forall \left(t_1,j_1,t_2,j_2,dist\right)
\label{zij}
\end{equation}

\begin{equation}
\sum_{c_{2} \in County | D\left(c_1,c_2\right) \neq d_{too far}}\sum_{t_{2} \in Type_{B}}\sum_{j_{2}\in Size_{B}} p^{x}_{t_1,j_1,t_2,j_2,D\left(c_1,c_2\right)} + dt^{x}_{c_1,t_1,j_1} + sl^{x}_{c_1,t_1,j_1} + st^{x}_{c_1,t_1,j_1} = 1.0 \text{     } \forall \left(c_1,t_1,j_1\right)
\label{outgoingdist}
\end{equation}
Inequality \ref{zij} serves to restrict the probabilities of movement, $p^{x}_{t_1,j_1,t_2,j_2,dist}$, to be less or equal to $1$ as $f_{t_1,j_1,t_2,j_2,dist}$ takes on a value of $1$ in the general case. By taking a value of $0$, it further prevents the movement of cattle from $Dairy$ subpopulations to $Preslaughter$ subpopulations ($t_{1} = Dairy$ and $t_{2} = Preslaughter$ $\Rightarrow$ $f_{t_1,j_1,t_2,j_2,dist} = 0$) and the out-going shipments of cattle from preslaughter subpopulations ($t_{1} = Preslaughter$ $\Rightarrow$ $f_{t_1,j_1,t_2,j_2,dist} = 0$). Equality constraint \ref{outgoingdist} ensures that the sum of the out-going probability distributions, the same distributions that are considered in the objective function, is equal to $1$.
\[
\text{Constraints on Movement data}
\]

\begin{IEEEeqnarray}{c}
\sum_{t_{1} \in Type_{B}}\sum_{j_{1}\in Size_{B}}\sum_{c_{2} \in County| D\left(c_1,c_2\right) \neq d_{too far}}\sum_{t_{2} \in Type_{B}}\sum_{j_{2}\in Size_{B}} Pop_{t_1,c_1,j_{1}}^{R}p^{x}_{t_1,j_1,t_2,j_2,D\left(c_1,c_2\right)}\nonumber\\
+ \sum_{t_{1} \in Type_{B}}\sum_{j_{1}\in Size_{B}} Pop_{t_1,c_1,j_{1}}^{R}sl^{x}_{c_1,t_1,j_1} + PN^{mov}\left(c_1\right) = \frac{Tc^{\left(s\right),x}_{All Movements,c_1}}{R_{C}} \text{     } \forall \left(c_1\right)
\label{totalsalesmove}
\end{IEEEeqnarray}

\begin{equation}
\sum_{j_{1}\in Size_{B}} Pop_{Preslaughter,c_1,j_{1}}^{R}sl^{x}_{c_1,Preslaughter,j_1} + PN^{slt}\left(c_1\right) = \frac{Tc^{\left(s\right),x}_{Slaughter,c_1}}{R_{C}} \text{     } \forall \left(c_1\right)
\label{totalsalesslaughter}
\end{equation}

\begin{equation}
Pop_{Preslaughter,c_1,z200\text{\_\itshape{up}}}^{R}sl^{x}_{c_1,Preslaughter,z200\text{\_\itshape{up}}} + PN^{slt500}\left(c_1\right) \geq \frac{Sales_{Slaughter,c_1,z500\text{\_\itshape{up}}}^{x}}{R_{C}} \text{     } \forall \left(c_1\right)
\label{largestsalesslaughter}
\end{equation}

\begin{equation}
D_{mov} = \sum_{c_{1} \in County}\left[\lvert PN^{mov}\left(c_1\right)\rvert + \lvert PN^{slt}\left(c_1\right)\rvert + PN^{slt500}\left(c_1\right) \right]
\label{errormove}
\end{equation}

Equality constraint \ref{totalsalesmove} sums the total sales and shipments originating in each county $c_{1}$ and tries to equate the total to the total sales and shipments defined by USDA NASS for county $c_{1}$, allowing a small amount of discrepancy through an error or roughness term $PN^{mov}\left(c_1\right)$. This discrepancy is permitted due to data challenges discussed in section \ref{datasection}. The scaling term, $R_{C} = 52.0$ weeks/year, converts the timescale of the estimation problem from a yearly to weekly basis for the estimated probabilities. Equality constraint \ref{totalsalesslaughter} equates the total slaughter from preslaughter feed subpopulations in each county $c_{1}$ to the respective, scaled data value from USDA NASS, again with a discrepancy term for each county $PN^{slt}\left(c_1\right)$. Inequality \ref{largestsalesslaughter} ensures that the largest yearly slaughter counts ($500$ or more head) are accredited to the largest ($200$ or more head) preslaughter subpopulations. This inequality requires a discrepancy term $PN^{slt500}\left(c_1\right)$ due to seasonality challenges in the USDA NASS data set. The discrepancy terms are collected in equality \ref{errormove}. Although we represent equality \ref{errormove} here with absolute value operators, the actual implementation linearizes the terms through a two-variable decomposition of the unrestricted variable that allows us to minimize the resulting value as if it were an absolute value \cite{T2007}. We retain the absolute value operators for simplicity in the formulation description.
\[
\text{Constraints on Population conservation}
\]

\begin{equation}
Leaving\left(c_{1},t_{1},j_{1},\right) = Pop_{t_1,c_1,j_{1}}^{R} \sum_{c_{2} \in County| D\left(c_1,c_2\right) \neq d_{too far}}\sum_{t_{2} \in Type_{B}}\sum_{j_{2}\in Size_{B}} p^{x}_{t_1,j_1,t_2,j_2,D\left(c_1,c_2\right)} \text{     } \forall \left(c_1,t_1,j_1\right)
\label{leaving}
\end{equation}

\begin{equation}
Coming\left(c_{1},t_{1},j_{1},\right) = \sum_{c_{2} \in County| D\left(c_2,c_1\right) \neq d_{too far}}\sum_{t_{2} \in Type_{B}}\sum_{j_{2}\in Size_{B}} Pop_{t_{2},c_{2},j_{2}}^{R} p^{x}_{t_2,j_2,t_1,j_1,D\left(c_2,c_1\right)} \text{     } \forall \left(c_1,t_1,j_1\right)
\label{coming}
\end{equation}

\begin{IEEEeqnarray}{c}
Leaving\left(c_{1},t_{1},j_{1},\right) - Coming\left(c_{1},t_{1},j_{1},\right) + \left(dt^{x}_{c_1,t_1,j_1} + sl^{x}_{c_1,t_1,j_1} - bt^{x}_{c_1,t_1,j_1}\right) Pop_{t_{1},c_{1},j_{1}}^{R} \nonumber\\
+ PN^{pop}\left(c_{1},t_{1},j_{1},\right) = 0.0 \text{     } \forall \left(c_1,t_1,j_1\right)
\label{popflux}
\end{IEEEeqnarray}

\begin{equation}
D_{pop} = \sum_{c_{1} \in County}\sum_{t_{1} \in Type_{B}}\sum_{j_{1}\in Size_{B}} \lvert PN^{pop}\left(c_{1},t_{1},j_{1},\right) \rvert
\label{errorpop}
\end{equation}

Equality constraints \ref{leaving} and \ref{coming} sum, respectively, the originating and arriving flows of cattle for each subpopulation of each county. Equality constraint \ref{popflux} then defines the total flux of every subpopulation in the system to be $0$ with small exceptions allowed through the discrepancy terms $PN^{pop}\left(c_{1},t_{1},j_{1},\right)$. Equality constraint \ref{errorpop} serves to aggregate the discrepancies. Here again in equality \ref{errorpop}, we retain the absolute value operator for simplicity in the formulation description \cite{T2007}.
\[
\text{Constraints on Industrial insights and discrepancies}
\]

\begin{equation}
D_{mov} + D_{pop} \leq f_{min}P_{All Cattle}^{\text{tot}}
\label{errorlimit}
\end{equation}

\begin{equation}
r^{expire-min}_{t_1} \leq dt^{x}_{c_1,t_1,j_1} \leq r^{expire-max}_{t_1} \text{     } \text{     } \forall \left(c_1,t_1,j_1\right)
\label{expirebounds}
\end{equation}

\begin{equation}
r^{slaughter-min}_{t_1} \leq sl^{x}_{c_1,t_1,j_1} \leq r^{slaughter-max}_{t_1} \text{     } \text{     } \forall \left(c_1,t_1,j_1\right)
\label{slaughterbounds}
\end{equation}

\begin{equation}
r^{birth-min}_{t_1} \leq bt^{x}_{c_1,t_1,j_1} \leq r^{birth-max}_{t_1} \text{     } \text{     } \forall \left(c_1,t_1,j_1\right)
\label{birthbounds}
\end{equation}

In inequality \ref{errorlimit}, we constrain the total discrepancies (counted in head of cattle) of the movements and net population fluxes to be less than a fraction $f_{min}$ of the total cattle in the system. The value of $f_{min}$ is determined by first solving the linear problem composed of the set of constraints of this formulation with an objective to minimize the system discrepancies. The value of $f_{min}$ is then taken as the ratio of the optimal objective value to the total system population and rounded up to the next highest thousandth. The inequality pairs \ref{expirebounds}, \ref{slaughterbounds}, and \ref{birthbounds} provide constraints by cattle type on the feasible probabilities used to describe the respective expiration, slaughter, and birth processes for each subpopulation.

\section{Optimization results} \label{resultssec}

We solved the cattle movement parameter problem of section \ref{formulation} using the AIMMS Modeling System of Paragon Decision Technology \cite{AIMMS2013}. The complete formulation is composed of $81142$ constraints with $80107$ variables and the objective function. For the error limit, a value of $f_{min} = 0.012$ was obtained through the method described following inequality \ref{errorlimit}, representing a limit of $1.2\%$ of the total number of cattle, $51,252,890$. The limits on the demographic probabilities attempt to capture loose bounds on feasible average rates of birth, culling, and slaughter. We assume that dairy cattle are not sent to slaughter houses through a slaughter rate $r^{slaughter-max}_{Dairy} = r^{slaughter-min}_{Dairy} = 0$, but rather through a culling process $r^{expire-min}_{Dairy} = (312 \text{ weeks})^{-1}, r^{expire-max}_{Dairy} = (104 \text{ weeks})^{-1}$. We bound the expected birthing rate of dairy cattle as $r^{birth-min}_{Dairy} = (62 \text{ weeks})^{-1}$ and $r^{birth-max}_{Dairy} = (36 \text{ weeks})^{-1}$. The mixed collection of cattle, $Beef$, are allowed a reasonable birth rate as well $r^{birth-max}_{Beef} = (52 \text{ weeks})^{-1}, r^{birth-min}_{Beef} = 0$, but the $Preslaughter$ individuals are not $r^{birth-min}_{Preslaughter} = r^{birth-max}_{Preslaughter} = 0$. The $Beef$ cattle have a maximum average useful lifespan defined by $r^{expire-min}_{Beef} = (520 \text{ weeks})^{-1}$ and they join the other two types in minimal useful life as $r^{expire-max}_{Beef} = r^{expire-max}_{Preslaughter} = (104 \text{ weeks})^{-1}$. The $Preslaughter$ population is assumed to not have a minimal expiration rate $r^{expire-min}_{Preslaughter} = 0$, but they have the highest feasible slaughter rate of $r^{slaughter-max}_{Preslaughter} = (2 \text{ weeks})^{-1}$. Lastly the $Beef$ populations have a feasible range for their slaughter rates of $r^{slaughter-min}_{Beef} = 0$ to $r^{slaughter-min}_{Beef} = (13 \text{ weeks})^{-1}$. The upper limits on the slaughter rates are quite high, but we explain the need for this later in this section.

\subsection{Cattle movement parameters}

\begin{figure}
	\includegraphics[width = 6.75in,  height = 2.67in]{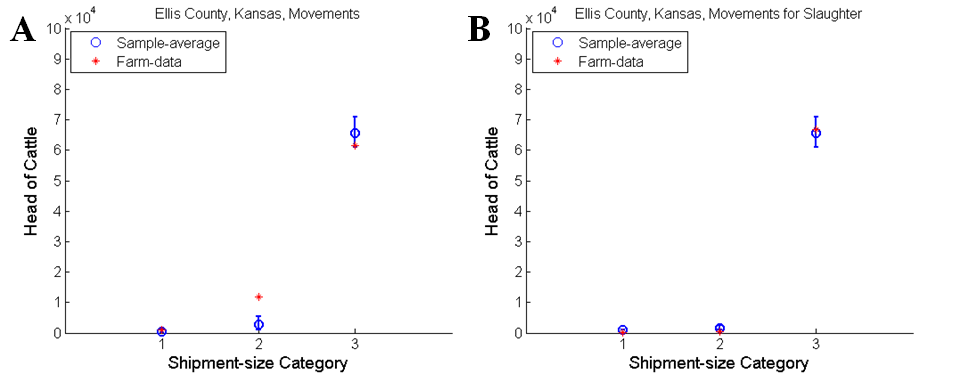}
	\caption[center]{(A) The average yearly totals of $All$ $Movements$ originating from Ellis County, Kansas and generated from the estimated subpopulation cattle movement parameters for the $3$ size categories of $Size_{B}$ are represented with blue circles and their respective $99\%$ confidence intervals are shown with the vertical lines. Shown in red stars are the yearly totals reported by NASS for the same county and aggregated into the ranges of $Size_{B}$. (B) The yearly totals of cattle shipped for slaughter originating in Ellis County are displayed from both the estimated slaughter rates (in blue) and the NASS database (in red) following the same notations as in A.}
	\label{EllisCounty}
\end{figure}

\begin{figure}
	\includegraphics[width = 6.75in,  height = 2.67in]{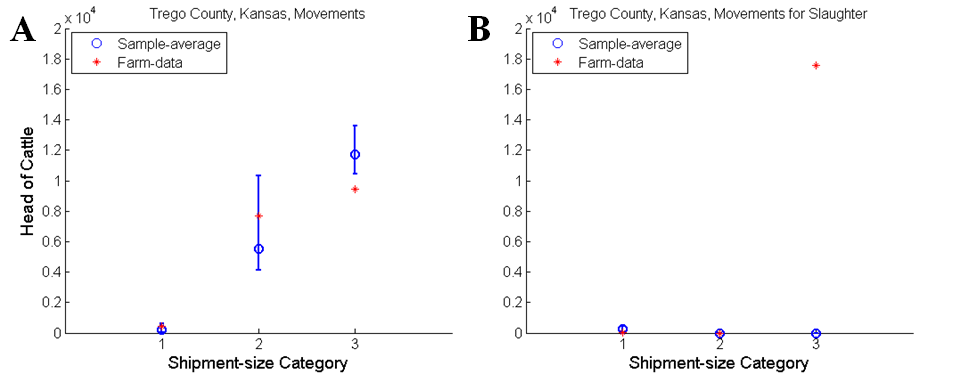}
	\caption[center]{(A) The average yearly totals of $All$ $Movements$ originating from Trego County, Kansas and generated from the estimated subpopulation cattle movement parameters for the $3$ size categories of $Size_{B}$ are represented with blue circles and their respective $99\%$ confidence intervals are shown with the vertical lines. Shown in red stars are the yearly totals reported by NASS for the same county and aggregated into the ranges of $Size_{B}$. (B) The yearly totals of cattle shipped for slaughter originating in Trego County are displayed from both the estimated slaughter rates (in blue) and the NASS database (in red) following the same notations as in A. A large discrepancy occurs in the third size category between the NASS reported slaughter totals and the generated distribution.}
	\label{TregoCounty}
\end{figure}

As the focus of this study, the cattle movement parameters $p^{x}_{t_1,j_1,t_2,j_2,dist}$ express the probability that, within a week's duration, an individual in a subpopulation of type $t_1$ and size $j_1$ will move or be shipped to a subpopulation of type $t_2$ and size $j_2$ at a (county-to-county) distance of $dist$. Table \ref{p_xTablepartial} presents a subset of these probabilities that express the movement of cattle from $Dairy$ subpopulations to $Beef$ subpopulations for $5$ ranges of distance.
\begin{longtable}{|l|c|c|c|c|c|}
\caption{Estimated cattle movement parameters $p^{x}_{t_1,j_1,t_2,j_2,dist} \cdot 10^{3}$, Dairy to Beef}\\
\hline
\textbf{Source} $\rightarrow$ \textbf{Destination} & $d0$ & $d100$ & $d200$ & $d500$ & $d1000$ \\
\hline
\endfirsthead
\multicolumn{6}{c}%
{\tablename\ \thetable\ -- \textit{Continued from previous page}} \\
\hline
\textbf{Source} $\rightarrow$ \textbf{Destination} & $d0$ & $d100$ & $d200$ & $d500$ & $d1000$ \\
\hline
\endhead
\hline \multicolumn{6}{r}{\textit{Continued on next page}} \\
\endfoot
\hline
\endlastfoot
    $D,z1\_19 \rightarrow B,z1\_19$ & 0.212302939 & 0.059831993 & 0.021823530 & 0.103473623 & 0.171239322 \\
    $D,z20\_199 \rightarrow B,z1\_19$ & 0.0 & 0.0 & 0.0 & 0.0 & 0.0 \\
		$D,z200\_up \rightarrow B,z1\_19$ & 0.043690097 & 0.007360407 & 0.000574583 & 0.0 & 0.0 \\
    $D,z1\_19 \rightarrow B,z20\_199$ & 0.184643994 & 0.054828320 & 0.021031163 & 0.108659569 & 0.275442544 \\
    $D,z20\_199 \rightarrow B,z20\_199$ & 0.0 & 0.0 & 0.0 & 0.0 & 0.0 \\
    $D,z200\_up \rightarrow B,z20\_199$ & 0.0 & 0.001677875 & 0.003638771 & 0.007653646 & 0.0 \\
		$D,z1\_19 \rightarrow B,z200\_up$ & 0.179577791 & 0.053431956 & 0.021227743 & 0.113325182 & 0.301730295 \\
		$D,z20\_199 \rightarrow B,z200\_up$ & 0.0 & 0.0 & 0.0 & 0.0 & 0.002141048 \\
    $D,z200\_up \rightarrow B,z200\_up$ & 0.0 & 0.001111461 & 0.004969657 & 0.019264109 & 0.001843747
  \label{p_xTablepartial}%
\end{longtable}%
A complete table of the cattle movement probabilities is provided in appendix B along with tables that partially present the birth, expiration, and slaughter probabilities of the $9$ subpopulations of each county. The entire results are too large for this document as we are studying $1034$ counties. The tables in appendix B present results for a sample of $10$ counties from each State. Once having obtained the solution, we revisited the movement data released by the USDA NASS. A significant difference exists between the NASS movement data and these movement parameters we've estimated, namely, that the movements of NASS are summarized from individual premises, but our parameters describe movements from and to collections of premises. We simulated $30$ years of virtual cattle movements and shipments for slaughter and summarized the movements into the $3$ size ranges of $Size_{B}$. We compare these results for individual counties, considering that shipments originating from aggregated subpopulations ought to usually be larger than shipments from individual operations. This means that the subpopulation-based results should over represent for larger sizes and perhaps under represent for smaller sizes of shipments. For Ellis County in the State of Kansas, figure \ref{EllisCounty} presents a comparison of our subpopulation based distributions of shipments in blue against the NASS reported yearly totals in red. The three size categories represent the three ranges of $Size_{B}$, with the smallest range on the left and the largest on the right side.
Trego County, also in the State of Kansas, demonstrates one way in which the year-long resolution of the Agricultural Census is insufficient to express the seasonality of the cattle system. At the time of the $2007$ census, Trego County reported no large $Preslaughter$ populations of cattle. On the year, however, Trego County was responsible for several large shipments of cattle for slaughter. The census happened to catch the finishing yards at a point in time in which they were empty and thus neither the true capacity nor typical population levels of $Preslaughter$ cattle are represented in the NASS database. Figure \ref{TregoCounty}.B displays the dramatic mismatch that occurs for the largest slaughter shipment size category.

\section{Discussion and conclusions} \label{concl}

We have designed and solved a large-scale optimal estimation problem in an attempt to address the privacy challenges and the need for livestock movement data in the United States. Given the resolution limitations of the data available, we do not try to estimate very detailed parameters, but rather adopt a stratified metapopulation approach where we shape the structure of our variables around the structure of the NASS data. Our approach is limited by the timescale resolution of the census report. This leads to several seasonal challenges that include correctly quantifying populations, identifying periods of higher and lower movement rates, and capturing birthing and slaughtering seasons. The demographic bounds used in our formulation depend on advice from industry experts and are flexibly open to further insights. We simulated and compared subpopulation-based movement distributions with operation-based movements, however this is not a rigorous method of validation. Our problem design includes an assumption on the relationships of cattle types and sizes that proved to not hold true for all counties as demonstrated by Broomfield County in the State of Colorado. Lastly, the parameters we estimate fail to consider individual State laws and Veterinary practices as well as the role of wildlife in the interfacing of subpopulations.\\

A close examination of the results in table B.1 of appendix B reveals a relatively unrealistic fraction of probabilities that are estimated to take on a value of $0$. We believe this to be an artifact that has arisen from the design of the formulation and the nature of the optimization. Without a sufficiently diverse set of constraints, the dimensionality of an optimal solution for the problem will be limited. The objective to maximize entropy would prefer to diversify the results, making as many nonzero as possible, yet it seems too tightly restricted in some way as to allow that to occur in the solution. If we had to select a constraint which would most likely be the cause for these probabilities taking on a value of $0$, we would first suspect the tight error limit of inequality \ref{errorlimit}. All other constraints allow a reasonable amount of flexibility in the set of parameters that would satisfy them. As the objective strives to diversify the distributions, we predict that loosening the error limits would result in fewer movement probabilities taking on a value of $0$.\\

From the challenges that this problem held, a few insights were discovered that might improve the success of a future version of these livestock movement estimation methods without posing any threat to the confidentiality of the data. Having examined the design limitations of this study, we would like to propose $3$ new questions to be considered for addition to future versions of the US Census of Agriculture. Those being
\begin{itemize}
  \item ``Of the cattle sold/moved in question $4$, excluding those sold for slaughter in question $6$, how many went to (a) destinations within the county, (b) destinations in a neighboring county, (c) destinations within the same state, (d) destinations in neighboring (bordering) states, (e) destinations further away?''
	\item ``Of cattle that arrive to this operation during $20$XX, how many came from (a) a locations within the county, (b) locations in a neighboring county, (c) locations within the same state, (d) locations in neighboring (bordering) states, (e) locations further away?''
	\item ``How many cattle were born on these premises during the year $20$XX?''
\end{itemize}
Although with many limitations, we have taken a significant first step in tackling the challenges of data in the United States through optimization and computation without compromising anonymity.\\

Large-scale livestock disease analysis and control is the most immediate use for the parameters estimated by this method. The threat of an livestock epidemic spreading across the region under consideration can be examined using a global epidemic invasion threshold for this system of cattle populations \cite{CV2007, BV2012}. A recent work has extended this global invasion threshold to directed livestock movements among many populations \cite{SSZB2014}. This work has derived two global epidemic invasion thresholds, where the first describes the emergence of the disease at a regional scale $\left(p_{c}\right)$ and the second describes the spreading of the disease into the preslaughter locations $\left(p_{c}^{TS}\right)$. We compute a rough approximation of the critical movement rates $p_{c}$ and $p_{c}^{TS}$ from the estimated data for the $10$ Central States. The computation of the average movement rate $ \left\langle p \right\rangle $ across the $10$ States and the comparison of these values to the critical movement rates reveal the potential for epidemics to break out across these States when considering the current movement rates. We use the same disease parameters for the theoretical comparison: infection rate $\beta = \mu R_{0}$, recovery rate $\mu = 1 (week)^{-1}$, and basic reproductive number $R_{0} = 1.2$. From our estimated parameters we determined the approximate average population size $\overline{N}^{approx} = 5483.8$ head and the average death rate $\delta = 0.01516 (week)^{-1}$. We simulated $30$ years of cattle movements and captured the weekly average outward movement $ \left\langle p \right\rangle $ for all $Beef$ and $Dairy$ subpopulations. Concurrently, we measured the resulting network parameters ($\eta_{in}$, $\eta_{out}$, $k_{i}^{in}$, $k_{i}^{out}$) for networks representing the weekly movements. We have considered the dynamic network measurements as produced by the realized movements. We have ignored the potential existence of any degree correlations in the resulting networks from the movement-based network construction. Having collected this approximate set of parameters, we computed the critical movement rates $p_{c}$ and $p_{c}^{TS}$ and list them in comparison to the system average movement rate $ \left\langle p \right\rangle $ of the estimated cattle movement parameters in table \ref{p_vs_pc_comp}.
\begin{longtable}{|l|c|c|c|}
\caption{Average cattle movement parameter and thresholds}\\
\hline
   & \textbf{Average} & \textbf{Minimum} & \textbf{Maximum} \\
\hline
\endfirsthead
\multicolumn{4}{c}%
{\tablename\ \thetable\ -- \textit{Continued from previous page}} \\
\hline
   & \textbf{Average} & \textbf{Minimum} & \textbf{Maximum} \\
\hline
\endhead
\hline \multicolumn{4}{r}{\textit{Continued on next page}} \\
\endfoot
\hline
\endlastfoot
    $ \left\langle p \right\rangle $ & 0.131892 & 0.128170 & 0.136130 \\
    $ p_{c} $ & 0.003470 & 0.003384 & 0.003565 \\
    $ p_{c}^{TS} $ & 0.007001 & 0.006745 & 0.007282 
  \label{p_vs_pc_comp}%
\end{longtable}%
This first approximation of the incorporation of these two components suggests that the US cattle systems are at a significant level of risk. The average cattle movement rates are roughly $2$ magnitudes larger than the thresholds defined by the critical movement rates. This result suggests that an epidemic with parameters similar to the ones implemented would easily invade US cattle populations and reach the cattle's final destinations, possibly compromising the security of the beef supply chain. To impede the progress of this epidemic, a reduction in cattle movement rates of more than $99\%$ would be required. For human influenza-like illnesses, a reduction in mobility rates of one order of magnitude has also been proposed \cite{CV2008}. This amount of reduction is based on an approximated comparison, and further work should be completed to better quantify the risk to US livestock systems.

\section{Acknowledgments}

The authors would like to express their gratitude to Dr. Sanderson, Dr. Schroeder, and local farmers and stockers for helpful insights into the beef cattle industry. This material is based upon work supported by the U. S. Department of Homeland Security under Grant Award Number $2010$-ST$061$-AG$0001$ and a grant through the Kansas Biosciences Authority. The views and conclusions contained in this publication are those of the authors and should not be interpreted as necessarily representing the official policies, either explicit or implicit, of the U. S. Department of Homeland Security and Kansas Biosciences Authority.

\section{Appendix A: Data Estimation}

Following the discussion of section \ref{datasection}, we formulated a pair of maximum entropy estimation problems that are constrained by straightforward rules defined by the US Department of Agriculture database. The maximum entropy method does not recover a significant amount of diversity with the estimated values, but selects the sets of values that are as homogeneous as possible while obeying all constraints. This method seeks such values because the homogeneity of maximum entropy represents the minimal assumptions possible in the problem design. 

\subsection{Population data estimation}

With the following formulations, we estimate any undisclosed data elements, namely subpopulations $Pop_{t,c,i}^{x}$, County totals $Tc_{t,c}^{x}$, and State totals for each size category $Tz_{t,i}^{x}$. We defined the sets $County$ as the Counties of the considered State, $t \in Type_{A}=\{Dairy$, $Preslaughter$, $All$ $Cattle\}$, $i \in Size_{A}=\{z1\_9$, $z10\_19$, $z20\_49$, $z50\_99$, $z100\_199$, $z200\_499$, $z500\_up\}$, and $j \in Size_{B}=\{z1\_19$, $z20\_199$, $z200\_up\}$. From the USDA NASS database, for each State, we collected the numbers of operations $n_{t,c,i}$ of cattle type $t$ having subpopulations with their sizes falling within size range $i$ in county $c$, the published populations counts $Pop_{t,c,i}^{r}$ representing all cattle of type $t$ with subpopulations within size range $i$ in county $c$, the total numbers of cattle $Tc_{t,c}^{r}$ of type $t$ in county $c$, the total numbers of cattle $Tz_{t,i}^{r}$ of type $t$ in size range $i$ in the State, and the total counts of cattle in each type $t$ for the entire State $P_{t}^{\text{tot}}$. To these parameters, we have added upper limits on the subpopulations $u_{t,i}$ for each cattle type $t$ and size range $i$, lower limits on the subpopulations $l_{t,i}$ for each cattle type $t$ and size range $i$, and data indicator parameters, $datPop_{t,c,i}$, $datTc_{t,c}$, and $datTz_{t,i}$, that express the existence of data elements for their respective parameters (The data indicators for non-disclosed data elements are assigned a value of $0$ and the remainder are set to $1$). The lower and upper limits are set by the limits of the size ranges ($l_{t,z1\_9} = 1$, $u_{t,z1\_9} = 9$, ...), with the exception of the upper limits on the largest size range, $z500\_up$. The largest upper limit $u_{t,z500\_up}$ is set to $Tz_{t,z500\_up}^{r}$ if $datTz_{t,z500\_up} = 1$, else we set $u_{t,z500\_up} = P_{t}^{\text{tot}}$. For the population data of each State, we solve
\begin{equation}
\text{Maximize   } \sum_{t \in Type_{A}}\sum_{c \in County}\sum_{i \in Size_{A}}\left( -\frac{Pop_{t,c,i}^{x}}{P_{t}^{\text{tot}}} \text{log}\left[\frac{Pop_{t,c,i}^{x}}{P_{t}^{\text{tot}}} + 1.0 - datPop_{t,c,i}\text{sign}\left(Pop_{t,c,i}^{r}\right)\right] \right) \text{ .}
\label{DE_Objective}
\end{equation}
The objective function maximizes the entropy of the three (by cattle type) population distributions and includes additional terms in the logarithm argument similar to the objective function of the formulation of section \ref{formulation}. We include the $sign$ function to consider the case when the published data $Pop_{t,c,i}^{r} = 0$, which would otherwise result in $log\left(0\right)$. Thus the $1.0$ is removed from the logarithm argument if and only if $datPop_{t,c,i} = 1$ and the corresponding data value $Pop_{t,c,i}^{r}$ is strictly positive.

\[
\text{Subject to}
\]

\[
\text{Constraints on Population values}
\]
\begin{equation}
Pop_{t,c,i}^{x} \le datPop_{t,c,i}Pop_{t,c,i}^{r} + \left(1.0 - datPop_{t,c,i}\right)n_{t,c,i}u_{t,i} \text{     } \forall \left(t, c, i\right)
\label{upperbound_Popx}
\end{equation}

\begin{equation}
Pop_{t,c,i}^{x} \ge datPop_{t,c,i}Pop_{t,c,i}^{r} + \left(1.0 - datPop_{t,c,i}\right)n_{t,c,i}l_{t,i} \text{     } \forall \left(t,c,i\right)
\label{lowerbound_Popx}
\end{equation}

The constraints form bounds for the complete set $\{Pop_{t,c,i}^{x}\}$ even if the data element $Pop_{t,c,i}^{r}$ is known from the NASS database. We use $datPop_{t,c,i}$ to represent the existence of the data element. Observe that when $datPop_{t,c,i} = 1$ inequalities \ref{upperbound_Popx} and \ref{lowerbound_Popx} converge to act as an equality constraint $Pop_{t,c,i}^{x}=Pop_{t,c,i}^{r}$. When $datPop_{t,c,i} = 0$, the upper (lower) bound is defined by the number of subpopulations in $Pop_{t,c,i}^{x}$ and the upper (lower) limit of the size range. Although these bounds have the potential to be very loose, they help shape the feasible set of values for $Pop_{t,c,i}^{x}$.

\[
\text{Constraints on County totals}
\]
\begin{equation}
Tc_{t,c}^{x} \le datTc_{t,c}Tc_{t,c}^{r} + \left(1.0 - datTc_{t,c}\right)uTc_{t,c} \text{     } \forall \left(t,c\right)
\label{upperbound_Tc}
\end{equation}

\begin{equation}
Tc_{t,c}^{x} \ge datTc_{t,c}Tc_{t,c}^{r} + \left(1.0 - datTc_{t,c}\right)lTc_{t,c} \text{     } \forall \left(t,c\right)
\label{lowerbound_Tc}
\end{equation}

\begin{equation}
uTc_{t,c} = \sum_{i \in Size_{A}}{\left[datPop_{t,c,i}Pop_{t,c,i}^{r} + \left(1.0 - datPop_{t,c,i}\right)n_{t,c,i}u_{t,i}\right]} \text{     } \forall \left(t,c\right)
\label{uTc}
\end{equation}

\begin{equation}
lTc_{t,c} = \sum_{i \in Size_{A}}{\left[datPop_{t,c,i}Pop_{t,c,i}^{r} + \left(1.0 - datPop_{t,c,i}\right)n_{t,c,i}l_{t,i}\right]} \text{     } \forall \left(t,c\right)
\label{lTc}
\end{equation}

\begin{equation}
\sum_{i \in Size_{A}}{Pop_{t,c,i}^{x}} = Tc_{t,c}^{x}\text{     } \forall \left(t,c\right)
\label{countytotals}
\end{equation}

Inequalities \ref{upperbound_Tc} and \ref{lowerbound_Tc} follow the same data indicator controlled constraint structure as inequalities \ref{upperbound_Popx} and \ref{lowerbound_Popx}. The upper and lower limits for the county totals $Tc_{t,c}^{x}$ are constructed in equality constraints \ref{uTc} and \ref{lTc}. Given constraints \ref{upperbound_Popx}, \ref{lowerbound_Popx}, and \ref{countytotals}, the bounds computed in \ref{uTc} and \ref{lTc} are redundant, however, we included them to help describe the solution space in a more understandable manner.

\[
\text{Constraints on Size totals}
\]
\begin{equation}
Tz_{t,i}^{x} \le datTz_{t,i}Tz_{t,i}^{r} + \left(1.0 - datTz_{t,i}\right)uTz_{t,i} \text{     } \forall \left(t,i\right)
\label{upperbound_Ti}
\end{equation}

\begin{equation}
Tz_{t,i}^{x} \ge datTz_{t,i}Tz_{t,i}^{r} + \left(1.0 - datTz_{t,i}\right)lTz_{t,i} \text{     } \forall \left(t,i\right)
\label{lowerbound_Ti}
\end{equation}

\begin{equation}
uTz_{t,i} = \sum_{c \in County}{\left[datPop_{t,c,i}Pop_{t,c,i}^{r} + \left(1.0 - datPop_{t,c,i}\right)n_{t,c,i}u_{t,i}\right]} \text{     } \forall \left(t,i\right)
\label{uTz}
\end{equation}

\begin{equation}
lTz_{t,i} = \sum_{c \in County}{\left[datPop_{t,c,i}Pop_{t,c,i}^{r} + \left(1.0 - datPop_{t,c,i}\right)n_{t,c,i}l_{t,i}\right]} \text{     } \forall \left(t,i\right)
\label{lTz}
\end{equation}

\begin{equation}
\sum_{c \in County}{Pop_{t,c,i}^{x}} = Tz_{t,i}^{x}\text{     } \forall \left(t,i\right)
\label{sizetotals}
\end{equation}

Constraints \ref{upperbound_Ti} through \ref{sizetotals} repeat the structure of constraints \ref{upperbound_Tc} through \ref{countytotals} for the totals $Tz_{t,i}^{x}$ of each size category.

\[
\text{Constraints on State totals}
\]

\begin{equation}
\sum_{c \in County}{Tc_{t,c}^{x}} = P_{t}^{\text{tot}} \text{     } \forall \left(t\right)
\label{sizesum}
\end{equation}

\begin{equation}
\sum_{i \in Size_{A}}{Tz_{t,i}^{x}} = P_{t}^{\text{tot}} \text{     } \forall \left(t\right)
\label{countysum}
\end{equation}

Equality constraints \ref{sizesum} and \ref{countysum} state that the totals of the sub-totals must equal the State total $P_{t}^{\text{tot}}$ of the respective population type.

\[
\text{Constraints on Population relations}
\]

\begin{equation}
Pop_{t,c,z1\text{\_}19}^{R} = Pop_{t,c,z1\text{\_}9}^{x} + Pop_{t,c,z10\text{\_}19}^{x}\text{     } \forall \left(t\right)
\label{smallfarms}
\end{equation}

\begin{equation}
Pop_{t,c,z20\text{\_}199}^{R} = Pop_{t,c,z20\text{\_}49}^{x} + Pop_{t,c,z50\text{\_}99}^{x} + Pop_{t,c,z100\text{\_}199}^{x}\text{     } \forall \left(t\right)
\label{mediumfarms}
\end{equation}

\begin{equation}
Pop_{t,c,z200\text{\_up}}^{R} = Pop_{t,c,z200\text{\_}499}^{x} + Pop_{t,c,z500\text{\_\itshape{up}}}^{x}\text{     } \forall \left(t\right)
\label{largefarms}
\end{equation}

\begin{equation}
Pop_{Dairy,c,j}^{R} + Pop_{Preslaughter,c,j}^{R} \le Pop_{All Cattle,c,j}^{R} \text{     } \forall \left(c \in County, j \in Size_{B}\right)
\label{sumtype}
\end{equation}

Constraints \ref{smallfarms}-\ref{largefarms} aggregate the populations from the size ranges of $Size_{A}$ to those of $Size_{B}$. Inequality \ref{sumtype} defines the relationship discussed in section \ref{datasection} and enables us to assemble the third cattle type of $Type_{B}$, $Beef$, as $Pop_{Beef,c,j}^{R} = Pop_{All Cattle,c,j}^{R} - Pop_{Dairy,c,j}^{R} - Pop_{Preslaughter,c,j}^{R}$ for every county $c$ and $j \in Type_{B}$.

\subsection{Sales and shipments data estimation}

For Sales, Movements, and Slaughter data, we consider another a set of shipment types $ShipType = \{All$ $Shipments$, $Slaughter\}$. We formulated the parallel problem for the non-disclosed data elements which describe the movements and slaughter of cattle. To estimate shipments $Sales_{q,c,i}^{x}$, County totals $Tc_{q,c}^{\left(s\right),x}$, and State totals for each size category $Tz_{q,i}^{\left(s\right),x}$ we solve
\begin{equation}
\text{Maximize   } \sum_{q \in ShipType}\sum_{c \in County}\sum_{i \in Size_{A}}\left( -\frac{Sales_{q,c,i}^{x}}{S_{q}^{\text{tot}}} \text{log}\left[\frac{Sales_{q,c,i}^{x}}{S_{q}^{\text{tot}}} + 1.0 - datSales_{q,c,i}\text{sign}\left(Sales_{q,c,i}^{r}\right)\right] \right) \text{ .}
\label{DE_ObjectiveSl}
\end{equation}

\[
\text{Subject to}
\]

\[
\text{Constraints on Population values}
\]
\begin{equation}
Sales_{q,c,i}^{x} \le datSales_{q,c,i}Sales_{q,c,i}^{r} + \left(1.0 - datSales_{q,c,i}\right)n^{\left(s\right)}_{q,c,i}u^{\left(s\right)}_{q,i} \text{     } \forall \left(q \in ShipType,c \in County,i \in Size_{A}\right)
\label{upperbound_PopxSl}
\end{equation}

\begin{equation}
Sales_{q,c,i}^{x} \ge datSales_{q,c,i}Sales_{q,c,i}^{r} + \left(1.0 - datSales_{q,c,i}\right)n^{\left(s\right)}_{q,c,i}l^{\left(s\right)}_{q,i} \text{     } \forall \left(q,c,i\right)
\label{lowerbound_PopxSl}
\end{equation}

\[
\text{Constraints on County totals}
\]
\begin{equation}
Tc^{\left(s\right),x}_{q,c} \le datTc^{\left(s\right)}_{q,c}Tc^{\left(s\right),r}_{q,c} + \left(1.0 - datTc^{\left(s\right)}_{q,c}\right)uTc^{\left(s\right)}_{q,c} \text{     } \forall \left(q,c\right)
\label{upperbound_TcSl}
\end{equation}

\begin{equation}
Tc^{\left(s\right),x}_{q,c} \ge datTc^{\left(s\right)}_{q,c}Tc^{\left(s\right),r}_{q,c} + \left(1.0 - datTc^{\left(s\right)}_{q,c}\right)lTc^{\left(s\right)}_{q,c} \text{     } \forall \left(q,c\right)
\label{lowerbound_TcSl}
\end{equation}

\begin{equation}
uTc^{\left(s\right)}_{q,c} = \sum_{i \in Size_{A}}{\left[datSales_{q,c,i}Sales_{q,c,i}^{r} + \left(1.0 - datSales_{q,c,i}\right)n^{\left(s\right)}_{q,c,i}u^{\left(s\right)}_{q,i}\right]} \text{     } \forall \left(q,c\right)
\label{uTcSl}
\end{equation}

\begin{equation}
lTc^{\left(s\right)}_{q,c} = \sum_{i \in Size_{A}}{\left[datSales_{q,c,i}Sales_{q,c,i}^{r} + \left(1.0 - datSales_{q,c,i}\right)n^{\left(s\right)}_{q,c,i}l^{\left(s\right)}_{q,i}\right]} \text{     } \forall \left(q,c\right)
\label{lTcSl}
\end{equation}

\begin{equation}
\sum_{i \in Size_{A}}{Sales_{q,c,i}^{x}} = Tc^{\left(s\right),x}_{q,c}\text{     } \forall \left(q,c\right)
\label{countytotalsSl}
\end{equation}

\[
\text{Constraints on Size totals}
\]
\begin{equation}
Tz_{q,i}^{\left(s\right),x} \le datTz_{q,i}^{\left(s\right)}Tz_{q,i}^{\left(s\right),r} + \left(1.0 - datTz^{\left(s\right)}_{q,i}\right)uTz^{\left(s\right)}_{q,i} \text{     } \forall \left(q,i\right)
\label{upperbound_TiSl}
\end{equation}

\begin{equation}
Tz_{q,i}^{\left(s\right),x} \ge datTz^{\left(s\right)}_{q,i}Tz_{q,i}^{\left(s\right),r} + \left(1.0 - datTz^{\left(s\right)}_{q,i}\right)lTz^{\left(s\right)}_{q,i} \text{     } \forall \left(q,i\right)
\label{lowerbound_TiSl}
\end{equation}

\begin{equation}
uTz^{\left(s\right)}_{q,i} = \sum_{c \in County}{\left[datSales_{q,c,i}Sales_{q,c,i}^{r} + \left(1.0 - datSales_{q,c,i}\right)n^{\left(s\right)}_{q,c,i}u^{\left(s\right)}_{q,i}\right]} \text{     } \forall \left(q,i\right)
\label{uTzSl}
\end{equation}

\begin{equation}
lTz^{\left(s\right)}_{q,i} = \sum_{c \in County}{\left[datSales_{q,c,i}Sales_{q,c,i}^{r} + \left(1.0 - datSales_{q,c,i}\right)n^{\left(s\right)}_{q,c,i}l^{\left(s\right)}_{q,i}\right]} \text{     } \forall \left(q,i\right)
\label{lTzSl}
\end{equation}

\begin{equation}
\sum_{c \in County}{Sales_{q,c,i}^{x}} = Tz_{q,i}^{\left(s\right),x}\text{     } \forall \left(q,i\right)
\label{sizetotalsSl}
\end{equation}

\[
\text{Constraints on State totals}
\]

\begin{equation}
\sum_{c \in County}{Tc_{q,c}^{\left(s\right),x}} = S_{q}^{\text{tot}} \text{     } \forall \left(q\right)
\label{sizesumSl}
\end{equation}

\begin{equation}
\sum_{i \in Size_{A}}{Tz_{q,i}^{\left(s\right),x}} = S_{q}^{\text{tot}} \text{     } \forall \left(q\right)
\label{countysumSl}
\end{equation}

\[
\text{Constraint on Shipment relations}
\]

\begin{equation}
Tc_{Slaughter,c}^{\left(s\right),x} \le Tc_{All Movement,c}^{\left(s\right),x} \text{     } \forall \left(c,i\right)
\label{sumtypeSl}
\end{equation}

The primary structural difference in this pair of optimization problems is the aggregations and relations of the last constraints of each. The shipment formulation only relates the two types through county totals $Tc_{q,c}^{\left(s\right),x}$ and not through the sub-elements as in the population formulation. For the objectives of this paper, there is not a need to add any assumptions to the relationship between the slaughter shipments and the total shipments beyond the assumption-less relationship of inequality \ref{sumtypeSl}. The aggregated populations $Pop_{t,c,j}^{R}$, the county shipment totals $Tc_{q,c}^{\left(s\right),x}$, and the largest category of slaughter shipments $Sales_{Slaughter,c,z500\_up}^{x}$ compose the set of inputs for the estimation of cattle movement parameters described in section \ref{formulation}.

\section{Appendix B: Results of Optimization}

In appendix A, we described the formulations of two optimization problems that estimate all non-disclosed elements of the particular USDA NASS data sets which are used in this chapter. We quantify the amount of estimated data for each State in tables \ref{PopEstTable} and \ref{MovEstTable}. Table \ref{PopEstTable} presents the numbers of populations estimated for each State by count and percentage as well as the number (head) of cattle assigned across these populations. The populations are counted by summing over all three types of $Type_{A} = \{Dairy$, $Preslaughter$, $All$ $Cattle\}$. This method of counting induces double counting since cattle in the first two types also belong to the third type, but offers a systematic quantification of the amount of data estimated. The percentage of cattle assigned through the estimation demonstrates that in situations where many populations are estimated the significance of these populations is less $< 7\%$. The States of Kansas and Texas are exceptions to the trend of small fractions of the total cattle being assigned through estimation. These two States appear to have relatively similar percentages of estimated populations when compared to the other States, yet they assign larger percentages of the cattle totals. These larger numbers of cattle suggest that a higher number of counties in these States possess only a few large cattle operations, where the sparsity and the size of the operations necessitate the non-disclosure of their data elements. Table \ref{MovEstTable} presents a parallel quantification for the estimations of the shipment distributions with counts similarly aggregated over $ShipType = \{All$ $Shipments$, $Slaughter\}$. The results of table \ref{MovEstTable} are comparable to those of table \ref{PopEstTable}.
\begin{longtable}{|l|c|c|c|c|}
\hline
\textbf{State} & \textbf{Count} & \textbf{Count} $\%$ & \textbf{Head} & \textbf{Head} $\%$ \\
\hline
\endfirsthead
\multicolumn{5}{c}%
{\tablename\ \thetable\ -- \textit{Continued from previous page}} \\
\hline
\textbf{State} & \textbf{Count} & \textbf{Count} $\%$ & \textbf{Head} & \textbf{Head} $\%$ \\
\hline
\endhead
\hline \multicolumn{5}{r}{\textit{Continued on next page}} \\
\endfoot
\hline
\caption{Totals and percentages of estimated cattle populations}
\endlastfoot
    Arkansas & $170$ & $10.79$ & $55,748$ & $3.06$ \\
    Colorado & $221$ & $16.44$ & $178,472$ & $4.46$ \\
    Iowa & $391$ & $18.81$ & $149,802$ & $2.52$ \\
    Kansas & $524$ & $23.76$ & $1,153,577$ & $12.20$ \\
    Minnesota & $387$ & $21.18$ & $132,701$ & $3.83$ \\
		Missouri & $500$ & $31.33$ & $112,753$ & $2.51$ \\
		Nebraska & $399$ & $20.43$ & $552,159$ & $5.89$ \\
		Oklahoma & $330$ & $20.41$ & $381,148$ & $6.55$ \\
		South Dakota & $282$ & $20.35$ & $112,648$ & $2.62$ \\
    Texas & $697$ & $13.07$ & $1,513,115$ & $8.81$ 
\label{PopEstTable}
\end{longtable}%

\begin{longtable}{|l|c|c|c|c|}
\hline
\textbf{State} & \textbf{Count} & \textbf{Count} $\%$ & \textbf{Head} & \textbf{Head} $\%$ \\
\hline
\endfirsthead
\multicolumn{5}{c}%
{\tablename\ \thetable\ -- \textit{Continued from previous page}} \\
\hline
\textbf{State} & \textbf{Count} & \textbf{Count} $\%$ & \textbf{Head} & \textbf{Head} $\%$ \\
\hline
\endhead
\hline \multicolumn{5}{r}{\textit{Continued on next page}} \\
\endfoot
\hline
\caption{Totals and percentages of estimated cattle shipments}
\endlastfoot
    Arkansas & $139$ & $13.24$ & $75,326$ & $7.56$ \\
    Colorado & $160$ & $17.86$ & $165,694$ & $3.20$ \\
    Iowa & $165$ & $11.90$ & $163,735$ & $2.75$ \\
    Kansas & $327$ & $22.24$ & $1,958,683$ & $13.71$ \\
    Minnesota & $182$ & $14.94$ & $108,631$ & $4.74$ \\
		Missouri & $334$ & $20.93$ & $213,464$ & $8.18$ \\
		Nebraska & $242$ & $15.16$ & $1,144,166$ & $8.98$ \\
		Oklahoma & $200$ & $18.55$ & $407,340$ & $8.74$ \\
		South Dakota & $156$ & $16.88$ & $155,108$ & $4.37$ \\
    Texas & $720$ & $20.25$ & $3,055,251$ & $17.38$ 
\label{MovEstTable}
\end{longtable}%

Table \ref{p_xTable} displays the solution for $p^{x}_{t_1,j_1,t_2,j_2,dist}$ sorted by the origin and destination pairs in the left column and the distances between the county centers in the rows. The cattle types of $Dairy$, $Beef$, and $Preslaughter$ are denoted respectively by $D$, $B$, and $P$ to be brief.
\begin{longtable}{|l|c|c|c|c|c|}
\caption{Estimated cattle movement parameters $p^{x}_{t_1,j_1,t_2,j_2,dist} \cdot 10^{3}$}\\
\hline
\textbf{Source} $\rightarrow$ \textbf{Destination} & $d0$ & $d100$ & $d200$ & $d500$ & $d1000$ \\
\hline
\endfirsthead
\multicolumn{6}{c}%
{\tablename\ \thetable\ -- \textit{Continued from previous page}} \\
\hline
\textbf{Source} $\rightarrow$ \textbf{Destination} & $d0$ & $d100$ & $d200$ & $d500$ & $d1000$ \\
\hline
\endhead
\hline \multicolumn{6}{r}{\textit{Continued on next page}} \\
\endfoot
\hline
\endlastfoot
    $D,z1\_19 \rightarrow D,z1\_19$ & 0.317036819 & 0.182356046 & 0.045722672 & 0.137186552 & 0.195279372 \\
    $D,z20\_199 \rightarrow D,z1\_19$ & 2.696121271 & 0.053763685 & 0.012881809 & 0.001997330 & 0.0 \\
    $D,z200\_up \rightarrow D,z1\_19$ & 0.327382526 & 0.005261100 & 0.002034078 & 0.006318463 & 0.005878289 \\
    $D,z1\_19 \rightarrow D,z20\_199$ & 0.033817329 & 0.000427105 & 0.000129380 & 0.001198734 & 0.009431513 \\
    $D,z20\_199 \rightarrow D,z20\_199$ & 0.0 & 0.0 & 0.0 & 0.0 & 0.0 \\
    $D,z200\_up \rightarrow D,z20\_199$ & 0.0 & 0.0 & 0.0 & 0.0 & 0.0 \\
    $D,z1\_19 \rightarrow D,z200\_up$ & 0.123132895 & 0.050114050 & 0.017433703 & 0.097523882 & 0.135399270 \\
    $D,z20\_199 \rightarrow D,z200\_up$ & 0.000725058 & 0.0 & 0.0 & 0.0 & 0.0 \\
\end{longtable}%

Tables \ref{slaughterTable}, \ref{deathTable}, and \ref{birthTable} display the solutions of the slaughter, expiration, and birth probabilities for the first $10$ alphabetical counties in each of the $10$ States. To receive an electronic copy of the full set of these results, or of the full solutions of the two formulations described in appendix A, please contact the corresponding authors to place your request.

\begin{longtable}{|l|c|c|c|c|}
\caption{Estimated cattle slaughter probabilities $sl^{x}_{c_1,t_1,j_1}$}\\
\hline
\textbf{County\_State} & \textbf{Type} & $z1\_19$ & $z20\_199$ & $z200\_up$ \\
\hline
\endfirsthead
\multicolumn{5}{c}%
{\tablename\ \thetable\ -- \textit{Continued from previous page}} \\
\hline
\textbf{County\_State} & \textbf{Type} & $z1\_19$ & $z20\_199$ & $z200\_up$ \\
\hline
\endhead
\hline \multicolumn{5}{r}{\textit{Continued on next page}} \\
\endfoot
\hline
\endlastfoot
    Allen\_Kansas & P     & 0.033571193 & 0.5   & 0.036699097 \\
          & D     & 0     & 0     & 0 \\
          & B     & 0.007507848 & 0.009829014 & 0.010425229 \\
\hline
    Anderson\_Kansas & P     & 0.023255814 & 0.108863578 & 0.149660755 \\
          & D     & 0     & 0     & 0 \\
          & B     & 0.006436844 & 0.010131312 & 0.011449758 \\
\hline
    Atchison\_Kansas & P     & 0.023255814 & 0.329251249 & 0.1943457 \\
          & D     & 0     & 0     & 0 \\
\end{longtable}

\begin{longtable}{|l|c|c|c|c|}
\caption{Estimated cattle culling probabilities $dt^{x}_{c_1,t_1,j_1}$}\\
\hline
\textbf{County\_State} & \textbf{Type} & $z1\_19$ & $z20\_199$ & $z200\_up$ \\
\hline
\endfirsthead
\multicolumn{5}{c}%
{\tablename\ \thetable\ -- \textit{Continued from previous page}} \\
\hline
\textbf{County\_State} & \textbf{Type} & $z1\_19$ & $z20\_199$ & $z200\_up$ \\
\hline
\endhead
\hline \multicolumn{5}{r}{\textit{Continued on next page}} \\
\endfoot
\hline
\endlastfoot
    Allen\_Kansas & P     & 0.009615385 & 0.009615385 & 0.009615385 \\
          & D     & 0.008628627 & 0.009615385 & 0.009615385 \\
          & B     & 0.009615385 & 0.009615385 & 0.009615385 \\
    \hline
    Anderson\_Kansas & P     & 0.005643688 & 0.009615385 & 0.009615385 \\
          & D     & 0.009615385 & 0.009615385 & 0.009615385 \\
          & B     & 0.009615385 & 0.009615385 & 0.009615385 \\
    \hline
    Atchison\_Kansas & P     & 0.009615385 & 0.009615385 & 0.009615385 \\
          & D     & 0.009615385 & 0.009615385 & 0.009615385 \\
\end{longtable}

\begin{longtable}{|l|c|c|c|c|}
\caption{Estimated cattle birth probabilities $bt^{x}_{c_1,t_1,j_1}$}\\
\hline
\textbf{County\_State} & \textbf{Type} & $z1\_19$ & $z20\_199$ & $z200\_up$ \\
\hline
\endfirsthead
\multicolumn{5}{c}%
{\tablename\ \thetable\ -- \textit{Continued from previous page}} \\
\hline
\textbf{County\_State} & \textbf{Type} & $z1\_19$ & $z20\_199$ & $z200\_up$ \\
\hline
\endhead
\hline \multicolumn{5}{r}{\textit{Continued on next page}} \\
\endfoot
\hline
\endlastfoot
    Allen\_Kansas & P     & 0     & 0     & 0 \\
          & D     & 0.027777778 & 0.016129032 & 0.017953942 \\
          & B     & 0.018276606 & 0.019230769 & 0.019230769 \\
    \hline
    Anderson\_Kansas & P     & 0     & 0     & 0 \\
          & D     & 0.025586227 & 0.016129032 & 0.018222653 \\
          & B     & 0.018107286 & 0.019230769 & 0.019230769 \\
    \hline
    Atchison\_Kansas & P     & 0     & 0     & 0 \\
          & D     & 0.016129032 & 0.016129032 & 0.018877302 \\
\end{longtable}

\bibliographystyle{IEEE}
\bibliography{sampleQ}

\begin{thebibliography}{10}

\bibitem{BMP2003}
Gary~W Brester, John~M Marsh, and Ronald~L Plain,
\newblock ``International red meat trade,''
\newblock {\em Veterinary Clinics of North America: Food Animal Practice}, vol.
  19, no. 2, pp. 493--518, July 2003.

\bibitem{FDA2001}
Neil~M. Ferguson, Christl~A. Donnelly, and Roy~M. Anderson,
\newblock ``The foot-and-mouth epidemic in great britain: Pattern of spread and
  impact of interventions,''
\newblock {\em Science}, vol. 292, no. 5519, pp. 1155--1160, May 2001.

\bibitem{KWSMCHCKWG2001}
Matt~J. Keeling, Mark E.~J. Woolhouse, Darren~J. Shaw, Louise Matthews, Margo
  Chase-Topping, Dan~T. Haydon, Stephen~J. Cornell, Jens Kappey, John
  Wilesmith, and Bryan~T. Grenfell,
\newblock ``Dynamics of the 2001 {UK} foot and mouth epidemic: Stochastic
  dispersal in a heterogeneous landscape,''
\newblock {\em Science}, vol. 294, no. 5543, pp. 813--817, Oct. 2001.

\bibitem{GKK2006}
D.~M. Green, I.~Z. Kiss, and R.~R. Kao,
\newblock ``Modelling the initial spread of foot-and-mouth disease through
  animal movements,''
\newblock {\em Proceedings of the Royal Society B: Biological Sciences}, vol.
  273, no. 1602, pp. 2729--2735, Nov. 2006.

\bibitem{KGJK2007}
Rowland~R Kao, Darren~M Green, Jethro Johnson, and Istvan~Z Kiss,
\newblock ``Disease dynamics over very different time-scales: foot-and-mouth
  disease and scrapie on the network of livestock movements in the {UK},''
\newblock {\em Journal of the Royal Society Interface}, vol. 4, no. 16, pp.
  907--916, Oct. 2007.

\bibitem{BBSC2012}
Paolo Bajardi, Alain Barrat, Lara Savini, and Vittoria Colizza,
\newblock ``Optimizing surveillance for livestock disease spreading through
  animal movements,''
\newblock {\em Journal of The Royal Society Interface}, vol. 9, no. 76, pp.
  2814--2825, Nov. 2012.

\bibitem{XCSS2013}
Ling Xue, Lee~W. Cohnstaedt, H.~Morgan Scott, and Caterina Scoglio,
\newblock ``A hierarchical network approach for modeling rift valley fever
  epidemics with applications in north america,''
\newblock {\em {PLoS} {ONE}}, vol. 8, no. 5, pp. e62049, May 2013.

\bibitem{CBBVV2007}
Vittoria Colizza, Alain Barrat, Marc Barthelemy, Alain-Jacques Valleron, and
  Alessandro Vespignani,
\newblock ``Modeling the worldwide spread of pandemic influenza: Baseline case
  and containment interventions,''
\newblock {\em {PLoS} Med}, vol. 4, no. 1, pp. e13, Jan. 2007.

\bibitem{CV2008}
Vittoria Colizza and Alessandro Vespignani,
\newblock ``Epidemic modeling in metapopulation systems with heterogeneous
  coupling pattern: Theory and simulations,''
\newblock {\em Journal of Theoretical Biology}, vol. 251, no. 3, pp. 450--467,
  Apr. 2008.

\bibitem{BCGHRV2009}
Duygu Balcan, Vittoria Colizza, Bruno Gonçalves, Hao Hu, José~J. Ramasco, and
  Alessandro Vespignani,
\newblock ``Multiscale mobility networks and the spatial spreading of
  infectious diseases,''
\newblock {\em Proceedings of the National Academy of Sciences}, vol. 106, pp.
  21484--21489, Dec. 2009.

\bibitem{MA1979}
Robert~M. May and Roy~M. Anderson,
\newblock ``Population biology of infectious diseases: Part {II},''
\newblock {\em Nature}, vol. 280, no. 5722, pp. 455--461, Aug. 1979.

\bibitem{PV2001}
Romualdo Pastor-Satorras and Alessandro Vespignani,
\newblock ``Epidemic spreading in scale-free networks,''
\newblock {\em Physical Review Letters}, vol. 86, no. 14, pp. 3200--3203, Apr.
  2001.

\bibitem{BBV2008}
Alain Barrat, Marc Barthélemy, and Alessandro Vespignani,
\newblock {\em Dynamical Processes on Complex Networks},
\newblock Cambridge University Press, Cambridge, {UK}, 2008.

\bibitem{V2012}
Alessandro Vespignani,
\newblock ``Modelling dynamical processes in complex socio-technical systems,''
\newblock {\em Nature Physics}, vol. 8, no. 1, pp. 32--39, Jan. 2012.

\bibitem{DSV2013}
F.~Darabi~Sahneh, C.~Scoglio, and P.~Van~Mieghem,
\newblock ``Generalized epidemic mean-field model for spreading processes over
  multilayer complex networks,''
\newblock {\em {IEEE/ACM} Transactions on Networking}, vol. PP, no. 99, pp. 1,
  2013.

\bibitem{US_CENSUS_2013}
``{US} census bureau,'' Oct. 2013.

\bibitem{US_AGCENSUS_2013}
``{USDA} - {NASS}, census of agriculture,'' Oct. 2013.

\bibitem{CV2007}
Vittoria Colizza and Alessandro Vespignani,
\newblock ``Invasion threshold in heterogeneous metapopulation networks,''
\newblock {\em Physical Review Letters}, vol. 99, no. 14, pp. 148701, Oct.
  2007.

\bibitem{BV2012}
Duygu Balcan and Alessandro Vespignani,
\newblock ``Invasion threshold in structured populations with recurrent
  mobility patterns,''
\newblock {\em Journal of theoretical biology}, vol. 293, pp. 87--100, Jan.
  2012.

\bibitem{EC2000}
European Council,
\newblock ``European parliament and european council (2000) regulation ({EC)}
  no. 1760/ 2000 of 17 july 2000 establishing a system for the identification
  and registration of bovine animals and regarding labeling of beef and beef
  products and repealing council regulation ({EC)} no. 820/97,''
\newblock Tech. {R}ep. L 204, Off. J. Eur. Communities, Aug. 2000.

\bibitem{W2005}
Cerian~R Webb,
\newblock ``Farm animal networks: unraveling the contact structure of the
  british sheep population,''
\newblock {\em Preventive veterinary medicine}, vol. 68, no. 1, pp. 3--17, Apr.
  2005.

\bibitem{KDGK2006}
{R.R} Kao, L~Danon, {D.M} Green, and {I.Z} Kiss,
\newblock ``Demographic structure and pathogen dynamics on the network of
  livestock movements in great britain,''
\newblock {\em Proceedings of the Royal Society B: Biological Sciences}, vol.
  273, no. 1597, pp. 1999--2007, Aug. 2006.

\bibitem{BTCMG2006}
M~Bigras-Poulin, R~A Thompson, M~Chriel, S~Mortensen, and M~Greiner,
\newblock ``Network analysis of danish cattle industry trade patterns as an
  evaluation of risk potential for disease spread,''
\newblock {\em Preventive Veterinary Medicine}, vol. 76, no. 1-2, pp. 11--39,
  Sept. 2006.

\bibitem{KGK2006}
Istvan~Z. Kiss, Darren~M. Green, and Rowland~R. Kao,
\newblock ``The network of sheep movements within great britain: network
  properties and their implications for infectious disease spread,''
\newblock {\em Journal of The Royal Society Interface}, vol. 3, no. 10, pp.
  669--677, Oct. 2006.

\bibitem{BBMG2007}
Michel Bigras-Poulin, Kristen Barfod, Sten Mortensen, and Matthias Greiner,
\newblock ``Relationship of trade patterns of the danish swine industry animal
  movements network to potential disease spread,''
\newblock {\em Preventive Veterinary Medicine}, vol. 80, no. 2–3, pp.
  143--165, July 2007.

\bibitem{REC2007}
{S.E} Robinson, {M.G} Everett, and {R.M} Christley,
\newblock ``Recent network evolution increases the potential for large
  epidemics in the british cattle population,''
\newblock {\em Journal of the Royal Society Interface}, vol. 4, no. 15, pp.
  669--674, Aug. 2007.

\bibitem{RC2007}
S~E Robinson and R~M Christley,
\newblock ``Exploring the role of auction markets in cattle movements within
  great britain,''
\newblock {\em Preventive Veterinary Medicine}, vol. 81, no. 1-3, pp. 21--37,
  Sept. 2007.

\bibitem{BNAL2008}
{Filipa M. Baptista}, {Telmo Nunes}, {Virgilio Almeida}, and {Armando Louza},
\newblock ``Cattle movements in portugal - an insight into the potential use of
  network analysis,''
\newblock {\em Revista Portuguesa de Ciencias Veterinarias}, vol. 107, no.
  565-566, pp. 35--40, 2008.

\bibitem{BKC2008}
{M.L.} Brennan, R.~Kemp, and {R.M.} Christley,
\newblock ``Direct and indirect contacts between cattle farms in north-west
  england,''
\newblock {\em Preventive Veterinary Medicine}, vol. 84, no. 3–4, pp.
  242--260, May 2008.

\bibitem{VK2009}
Matthew~C. Vernon and Matt~J. Keeling,
\newblock ``Representing the {UK's} cattle herd as static and dynamic
  networks,''
\newblock {\em Proceedings of the Royal Society B: Biological Sciences}, vol.
  276, no. 1656, pp. 469--476, Feb. 2009.

\bibitem{NGSPPFC2009}
Fabrizio Natale, Armando Giovannini, Lara Savini, Diana Palma, Luigi Possenti,
  Gianluca Fiore, and Paolo Calistri,
\newblock ``Network analysis of italian cattle trade patterns and evaluation of
  risks for potential disease spread,''
\newblock {\em Preventive Veterinary Medicine}, vol. 92, no. 4, pp. 341--350,
  Dec. 2009.

\bibitem{MAF2009}
{{MAF} Biosecurity New Zealand},
\newblock ``Review of selected cattle identification and tracing systems
  worldwide,'' 2009,
\newblock 2009/03.

\bibitem{KDVH2010}
Matt~J. Keeling, Leon Danon, Matthew~C. Vernon, and Thomas~A. House,
\newblock ``Individual identity and movement networks for disease
  metapopulations,''
\newblock {\em Proceedings of the National Academy of Sciences}, vol. 107, no.
  19, pp. 8866--8870, May 2010.

\bibitem{VHSW2010}
Victoriya~V. Volkova, Richard Howey, Nicholas~J. Savill, and Mark E.~J.
  Woolhouse,
\newblock ``Sheep movement networks and the transmission of infectious
  diseases,''
\newblock {\em {PLoS} {ONE}}, vol. 5, no. 6, pp. e11185, June 2010.

\bibitem{RDD2011}
S~Rautureau, B~Dufour, and B~Durand,
\newblock ``Vulnerability of animal trade networks to the spread of infectious
  diseases: A methodological approach applied to evaluation and emergency
  control strategies in cattle, france, 2005,''
\newblock {\em Transboundary and Emerging Diseases}, vol. 58, no. 2, pp.
  110--120, Apr. 2011.

\bibitem{NAIS2013}
``National animal identification system - {US} department of agriculture,''
  Oct. 2013.

\bibitem{BBT1995}
{DeeVon} Bailey, B.~Wade Brorsen, and Michael~R. Thomsen,
\newblock ``Identifying buyer market areas and the impact of buyer
  concentration in feeder cattle markets using mapping and spatial
  statistics,''
\newblock {\em American Journal of Agricultural Economics}, vol. 77, no. 2, pp.
  309--318, May 1995.

\bibitem{FHDW1998}
{Forde K.}, {Hillberg-Seitzinger A.}, {Dargatz D.}, and {Wineland N.},
\newblock ``The availability of state-level data on interstate cattle movements
  in the united states,''
\newblock {\em Preventive Veterinary Medicine}, vol. 37, no. 1, pp. 209--217,
  1998.

\bibitem{BTC2001}
T~W Bates, M~C Thurmond, and T~E Carpenter,
\newblock ``Direct and indirect contact rates among beef, dairy, goat, sheep,
  and swine herds in three california counties, with reference to control of
  potential foot-and-mouth disease transmission,''
\newblock {\em American journal of veterinary research}, vol. 62, no. 7, pp.
  1121--1129, July 2001.

\bibitem{SM2007}
{D. Shields} and {K. Mathews},
\newblock ``2003 interstate livestock movements {USDA} {ERA} outlook report
  {LDP-M-108-01},''
\newblock Tech. {R}ep., {USDA} National Agricultural Statistics Service,
  {USDA}, 2007.

\bibitem{D2007}
{B. J. Dominguez},
\newblock {\em Characterization of livestock herds in extensive agricultural
  settings in southwest Texas},
\newblock {MS}, Texas {A\&M} University, {USA}, 2007.

\bibitem{THBCOASK2009}
Michael~J. Tildesley, Thomas~A. House, Mark~C. Bruhn, Ross~J. Curry, Maggie
  {O’Neil}, Justine L.~E. Allpress, Gary Smith, and Matt~J. Keeling,
\newblock ``Impact of spatial clustering on disease transmission and optimal
  control,''
\newblock {\em Proceedings of the National Academy of Sciences}, vol. 107, no.
  3, pp. 1041--1046, Dec. 2009.

\bibitem{BGPMLW2013}
Michael~G. Buhnerkempe, Daniel~A. Grear, Katie Portacci, Ryan~S. Miller,
  Jason~E. Lombard, and Colleen~T. Webb,
\newblock ``A national-scale picture of {U.S.} cattle movements obtained from
  interstate certificate of veterinary inspection data,''
\newblock {\em Preventive Veterinary Medicine}, , no. In Press, 2013.

\bibitem{NASS2013}
``National agricultural statistics service: {NASS},'' Oct. 2013.

\bibitem{LGHDRF1997}
Willard~C Losinger, Lindsey~P Garber, George~W Hill, Stephen~E Dornseif,
  Judith~M Rodriguez, and William~B Frye,
\newblock ``Design and implementation of the united states national animal
  health monitoring system 1994–1995 cattle on feed evaluation, and an
  evaluation of the impact of response biases,''
\newblock {\em Preventive Veterinary Medicine}, vol. 31, no. 1–2, pp. 1--14,
  July 1997.

\bibitem{FU2003}
Dillon~M Feuz and Wendy~J Umberger,
\newblock ``Beef cow-calf production,''
\newblock {\em Veterinary Clinics of North America: Food Animal Practice}, vol.
  19, no. 2, pp. 339--363, July 2003.

\bibitem{VC2009}
{Tom Vilsack} and {Cynthia {Z.F.} Clark},
\newblock ``2007 census of agriculture, united states, summary and state
  data,''
\newblock Tech. {R}ep. Volume 1, Part 51, {AC-07-A-51}, United States
  Department of Agriculture, Dec. 2009.

\bibitem{DDM2004}
David~A Dargatz, Grant~A Dewell, and Robert~G Mortimer,
\newblock ``Calving and calving management of beef cows and heifers on
  cow–calf operations in the united states,''
\newblock {\em Theriogenology}, vol. 61, no. 6, pp. 997--1007, Apr. 2004.

\bibitem{Y2008}
Raymond~W. Yeung,
\newblock {\em Information Theory and Network Coding},
\newblock Springer, Aug. 2008.

\bibitem{T2007}
Hamdy~A. Taha,
\newblock {\em Operations Research: An Introduction},
\newblock Prentice Hall International, 2007.

\bibitem{AIMMS2013}
``{AIMMS:} the modeling system,'' Oct. 2013.

\bibitem{SSZB2014}
P~Schumm, C~Scoglio, Q~Zhang, and D~Balcan,
\newblock ``Global epidemic invasion thresholds in directed subpopulation
  networks having source, sink, and transit nodes,''
\newblock {\em Under Review}, 2014.

\end{thebibliography}
\end{document}